\documentclass[apjl]{emulateapj}
\usepackage{enumerate}
\usepackage {epsfig}
\usepackage {verbatim}
\usepackage {amsmath}
\usepackage {graphicx}
\usepackage {wrapfig}
\shorttitle{VHE Gamma-Ray Cascade Model}
\shortauthors{H. Huan et al}

\begin{document}
\title{A New Model for Gamma-Ray Cascades in Extragalactic Magnetic Fields}
\author{H.~Huan\altaffilmark{1},T.~Weisgarber\altaffilmark{1},T.~Arlen\altaffilmark{2},S.~P.~Wakely\altaffilmark{1}}
\altaffiltext{1}{Enrico Fermi Institute, University of Chicago, Chicago, IL 60637, USA}
\altaffiltext{2}{Department of Physics and Astronomy, University of California, Los Angeles, CA 90095, USA}
\begin{abstract}
	Very-high-energy (VHE, $E\gtrsim100$ GeV) gamma rays emitted by extragalactic sources, such as blazars, initiate electromagnetic cascades in the intergalactic medium. The cascade
	photons arrive at the earth with angular and temporal distributions correlated with the extragalactic magnetic field (EGMF).
	We have developed a new semi-analytical model of the cascade
	properties which is more accurate than previous analytic approaches
	and faster than full Monte Carlo simulations. Within its range of applicability, our model can quickly generate cascade spectra for a variety of source
	emission models, EGMF strengths, and assumptions about the source
	livetime.
	In this Letter, we describe the properties of the model and demonstrate
	its utility by exploring the gamma-ray emission from the blazar RGB
	J0710+591. In particular, we predict, under various scenarios, the VHE
	and high-energy (HE, 100 MeV $\lesssim$ E $\lesssim$ 300 GeV) fluxes
	detectable with the VERITAS and \textit{Fermi} Large Area Telescope
	(LAT) observatories. We then develop a systematic framework for
	comparing the predictions to published results, obtaining constraints
	on the EGMF strength. At a confidence level of 95\%, we find the lower
	limit on the EGMF strength to be $\sim2\times 10^{-16}$ Gauss if no limit
	is placed on the livetime of the source or $\sim3\times 10^{-18}$ Gauss if
	the source livetime is limited to the past $\sim$ 3 years during which
	\textit{Fermi} observations have taken place.
\end{abstract}
\keywords{astroparticle physics --- BL Lacertae objects: individual (RGB J0710+591) --- cosmic background radiation --- gamma rays: general --- intergalactic medium --- magnetic fields}
\maketitle


\section{Introduction}\label{sec:intro}
	The extragalactic magnetic field (EGMF) is of great interest to the overall understanding of astrophysical magnetic fields and related processes. It could act
	as a seeding field for magnetic fields in galaxies and clusters~\citep{Widrow2002}, and its
	origin may be related to inflation or other periods in the early history of the universe~\citep{Grasso2001}. Faraday rotation measurements~\citep{Kronberg1982,
	Kronberg1994,Blasi1999} and analysis of COBE data anisotropy~\citep{Barrow1997,Durrer2000} have put an upper bound on the EGMF strength at $\sim10^{-9}$ Gauss.
	On the other hand, gamma-ray-initiated electromagnetic cascades deflected by the EGMF in intergalactic space have a characteristic angular
	spread~\citep{Aharonian1994} and time delay~\citep{Plaga1995}, both of which provide a probe for lower EGMF strengths~\citep{Neronov2007,Elyiv2009}.
	Present and next generation gamma-ray telescopes have the possibility to measure the EGMF
	strength by observing the angular and temporal distributions of cascade photons from extragalactic gamma-ray sources such as blazars~\citep{Neronov2009,Dolag2009}.

	With current very-high-energy (VHE, $E\gtrsim100$ GeV) and high-energy (HE, 100 MeV $\lesssim E\lesssim 300$ GeV) gamma-ray data on VHE-selected blazars, a lower limit on the EGMF strength can be placed by requiring that
	the cascade flux of VHE emission not exceed the measured flux or upper limit in the HE band~\citep{Murase2008,Neronov2010}. Analytic cascade models assuming a simple
	relationship between the cascade flux and EGMF strength have put the lower limit at $10^{-16}$ to $10^{-15}$ Gauss when the source livetime is
	unlimited~\citep{Tavecchio2010,Tavecchio2011,Neronov2010}, similar to the results of Monte Carlo simulations~\citep{Dolag2011,Taylor2011}. If the cascade time delay is limited to the $\sim3$ years of simultaneous HE and VHE observations, the lower limit becomes $10^{-19}$ to $10^{-18}$ Gauss according to the simple cascade models~\citep{Dermer2011}, or $10^{-18}$ to $10^{-17}$ Gauss according to the simulations~\citep{Taylor2011}.

	In this Letter, we present a new semi-analytic model of the electromagnetic cascade. In contrast to previous analytic cascade models~\citep{Dermer2011,Neronov2009,Tavecchio2011}, our cascade model considers
	the full track of the primary photon without the assumption of interacting exclusively at the mean free path. This simultaneously accounts for both angular and temporal constraints
	in a natural way. In addition, we model the radiation backgrounds and source emission in greater detail. As a complementary approach to full Monte Carlo simulations~\citep{Dolag2011,Taylor2011,preparation},
	our model serves as a tool for clarifying the cascade picture, rapidly searching through the parameter space, and interpreting simulation results.
	We also develop a systematic framework for applying our cascade model's predictions to derive lower limits on the EGMF strength at specific confidence levels. This framework is applicable to the results of Monte Carlo simulations as well.

\section{Mathematical Model of Cascades} \label{sec:model}
\begin {figure}[t]
		\begin {center}
			\includegraphics[width=\columnwidth,clip,trim=0 0 0 2.4in]{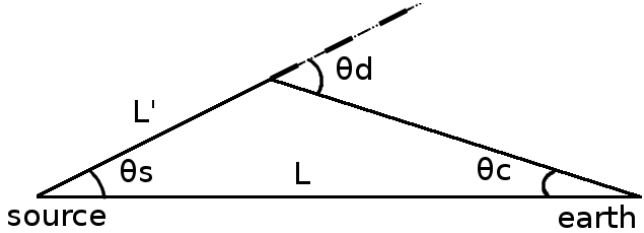}
			\caption{Geometry used to calculate the gamma-ray cascade, following~\cite{Dermer2011}.}
			\label{fig:geometry}
		\end {center}
	\end {figure}

	Consider a blazar emitting gamma rays at distance $L$.
	Gamma rays emitted at an angle $\theta_s$ relative to the line of sight may produce an $e^\pm$ pair via absorption on the photon
	background~\citep{Gould1967} at a distance $L'$ from the source, not necessarily equal to the mean free path. After being deflected by the EGMF
	through the angle $\theta_d$, the pairs could scatter background photons to gamma-ray energies, redirecting them toward the observer. These secondary gamma rays would
	arrive at an incidence angle $\theta_c$ (see Fig.~\ref{fig:geometry}). In this picture, the influence of the EGMF enters solely through $\theta_d$. The angles $\theta_c$ and $\theta_s$ are uniquely specified in terms of $\theta_d$, $L$, and $L'$, provided $\theta_c<\pi/2$.

	Because the energy density of CMB photons far exceeds that of the extragalactic background light (EBL), we assume that inverse Compton scattering proceeds in
	the Thomson regime via $e^\pm$ interactions with the CMB exclusively. An electron with Lorentz factor $\gamma_e$ will on average scatter secondary photons to
	energy $4\gamma_e^2\epsilon_0/3$, where $\epsilon_0\approx0.64$ meV is the average CMB photon energy~\citep{Blumenthal1970}. The energy loss rate of the
	electron is
	\begin {equation}\label{eq:loss}
		\frac{d\gamma_em_ec^2}{dt}=-\frac{4}{3}\epsilon_0cn_\textnormal{CMB}
		\sigma_T\gamma_e^2,
	\end {equation}
	where $c$ is the speed of light, $n_\textnormal{CMB}\approx411\text{ cm}^{-3}$ is the CMB photon density, and $\sigma_T\approx6.65\times10^{-25}\text{ cm}^2$
	is the Thomson cross section. For an EGMF $\textbf{\textit{B}}$ perpendicular to the electron momentum $\textbf{\textit{p}}_e$, the deflection rate in terms of
	the Larmor radius $r_l=\gamma_em_ec^2/eB$ is
	\begin {equation}\label{eq:deflection}
		\frac{d\theta_d}{dt}=\frac{c}{r_l}=\frac{eB}{\gamma_em_ec}.
	\end {equation}
	Thus, when an electron's Lorentz factor has changed from $\gamma_{e0}$ to $\gamma_e$, the electron will have been deflected by
	\begin {equation}\label{eq:defangle}
		\theta_{d0}(\gamma_{e0},\gamma_e)=\frac{3}{8}\frac{eB}{\epsilon_0
		n_\textnormal{CMB}\sigma_T}\left(\frac{1}{\gamma_e^2}-\frac{1}{\gamma_{e0}
		^2}\right).
	\end {equation}
	However, if the angle $\theta_f$ between $\textbf{\textit{B}}$ and $\textbf{\textit{p}}_e$ is other than $\pi/2$, Eq.~\ref{eq:defangle} should be generalized
	to
	\begin {equation}\label{eq:truedefangle}
		\theta_d(\gamma_{e0},\gamma_e,\theta_f)=\arccos\left(\sin^2\theta_f
		\cos\theta_{d0}+\cos^2\theta_f\right).
	\end {equation}
	
	We use the full CMB black-body spectrum to determine the observed cascade spectrum. The number of secondary photons between energies $E$ and $E+dE$ produced
	by an electron that changes Lorentz factor from $\gamma_e+d\gamma_e$ to $\gamma_e$ is the number density of CMB photons between $3E/(4\gamma_e^2)$ and $3(E+dE)/
	(4\gamma_e^2)$ times the Thomson cross section and the distance travelled by the electron:
	\begin {eqnarray}\label{eq:diffspec}
		dN(E,\gamma_e)&=&cdt\sigma_T\frac{27\pi E^2}
		{8\gamma_e^4}\frac{dE}{h^3c^3\gamma_e^2(e^{3E/4(\gamma_e^2kT)}-1)}\nonumber\\
		&=&\frac{81\pi E^2m_ed\gamma_e}{32h^3c\gamma_e^8
		\epsilon_0n_\textnormal{CMB}}\frac{dE}{e^{3E/(4\gamma_e^2kT)}-1}
	\end {eqnarray}
	where $h$ is the Planck constant, $k$ the Boltzmann constant, and $T=2.73$ K the CMB temperature.

	In terms of the mean free path $\lambda(\epsilon)=L/\tau(\epsilon)$ inferred from the optical depth $\tau(\epsilon)$ of the EBL, the probability for a photon of energy $\epsilon$ to be absorbed between $L'$ and $L'+dL'$ is
	$e^{-L'/\lambda(\epsilon)}dL'/\lambda(\epsilon)$. Approximating both particles in the resulting pair to have initial energy $m_e\gamma_{e0}c^2=\epsilon/2$, we calculate the differential flux
	of observed secondary photons by integrating over $L'$, $\epsilon$, and $\gamma_e$, and averaging over $\theta_f$:
	\begin {eqnarray}
		& &\frac{dN(E)}{dE}=\int \frac{d\gamma_e81\pi E^2m_e}{16h^3c
		\gamma_e^8\epsilon_0n_{CMB}\left(e^{3E/(4\gamma_e^2kT)}-1\right
		)}\nonumber\\
		&\times&\int d\theta_fg(\theta_f)\int d\epsilon\int dL'\frac{e^{-L'/\lambda(\epsilon)}}{\lambda
		(\epsilon)}f\left(\epsilon,\theta_s\right)\nonumber\\
		&\times& \exp\left(-\sqrt{L^2+L'^2-2LL'\cos(\theta_d-
		\theta_c)}/\lambda(E)\right)\label{eq:model}
	\end {eqnarray}
	Here, $g(\theta_f)$ is the probability distribution of $\theta_f$, equal to $\sin\theta_f$ for an EGMF uniformly distributed in direction, and
	$f(\epsilon,\theta_s)$ is the intrinsic flux of the source, with
	$\theta_s=\theta_d-\theta_c$ from Fig.~\ref{fig:geometry}. The $1/2\pi$ factor from the $e^{\pm}$ being deflected into the surface of a cone with opening angle
	$\theta_d$ cancels the $2\pi$ enhancement from a similar effect at the source. We take the integral over $\theta_f$ from $0$ to $\pi/2$. The physical lower bound
	for the integration over the primary energy $\epsilon$ is $\epsilon_\textnormal{min}=2\gamma_em_ec^2$. Nearly all of the photons above 200 TeV will be absorbed within
	1 Mpc of the source, and the $e^\pm$ pairs will be isotropized by the strong field of the surrounding galaxy, resulting in negligible cascade contribution. We
	therefore adopt an upper limit of $\epsilon_\textnormal{max}=200$ TeV, suggesting an upper limit of $\epsilon_\textnormal{max}/2m_ec^2$ on the $\gamma_e$
	integration. As a practical matter, we enforce a lower limit of $\gamma_e=10^5$ on the $\gamma_e$ integration, motivated by the CMB density becoming
	negligible at energies above 3 meV and our disinterest in the cascade spectrum below 100 MeV.

	Observational effects enter Eq.~\ref{eq:model} through limits on the $L'$ integration. As seen in Fig.~\ref{fig:geometry}, we can express $\theta_c$ as
	\begin {equation}\label{eq:angle}
		\theta_c=\arcsin\left(\frac{L'}{L}\sin\theta_d\right)
	\end {equation}
	so that a cut on $\theta_c$ translates directly into a cut on $L'$. Similarly, the time delay $\Delta T$ of cascade photons may be written as
	\begin {equation}\label{eq:time}
		c\Delta T=L'+\sqrt{L^2+L'^2-2LL'\cos(\theta_d-
		\theta_c)}-L,
	\end {equation}
	exchanging a limit on the source livetime $\Delta T$ for another constraint on $L'$. We adopt the intersection of the $L'$ cuts from Eqs.~\ref{eq:angle} and~\ref{eq:time} in evaluating  the cascade flux via Eq.~\ref{eq:model}.

	We now briefly examine several assumptions made in the construction of this cascade model.
	(i) Exact energy distributions of pair production products are approximated as each having half the energy of the primary photon, as the cascade spectrum only
	weakly depends on the spectral distribution of pairs~\citep{Coppi1997}.
	(ii) The Thomson limit assumption, appropriate when $\sqrt{\epsilon_0 E}\ll m_ec^2$ $\Rightarrow$ $E\ll400$ TeV, certainly produces spectra valid for the range of existing TeV gamma-ray instruments.
	(iii) We assume the EGMF to be coherent over the cooling length of the electrons, at most a few Mpc for the most energetic electrons, making this assumption valid for coherence lengths $\lambda_B\gtrsim$ 1 Mpc.
	(iv) Only secondary cascade photons are considered.
	For observed photons above 100 MeV, the primary energy must be $\epsilon\gtrsim 2m_ec^2\sqrt{3\times (100\text{ MeV})/4\epsilon_0}\approx400$ GeV. The requirement for third-generation photons is thus $\epsilon\gtrsim 2m_ec^2\sqrt{3\times (400\text{ GeV})/4\epsilon_0}\approx25$ TeV.
	While a primary photon above this energy does produce higher-generation cascade photons, the power going into the higher generations is small for conventional
	blazar emission models, leading to negligible contribution from higher-order cascades~\citep{Tavecchio2011}.
	(v) Cosmological effects enter the cascade model solely in the calculation of $\lambda(\epsilon)$. Cosmic expansion, energy redshift, EBL evolution, and other cosmological effects are ignored, limiting the application of the cascade model to nearby sources ($z\lesssim0.2$), considering the $(1+z)^4$ radiation density evolution.
	(vi) We assume axial symmetry in the intrinsic emission $f(\epsilon,\theta_s)$, following the current understanding of blazar emission, that is, boosted isotropic emission from hot blobs with the jet direction pointing toward the earth~\citep{Urry1995}.
	(vii) The blazar intrinsic emission should be steady over the time interval $\Delta T$.

	To get the predicted cascade flux from Eq.~\ref{eq:model}, we require a source emission model $f(\epsilon,\theta_s)$. Motivated by models of blazars as
	relativistically beamed sources~\citep{Urry1995} we model the intrinsic emission as boosted isotropic emission,
	\begin {eqnarray}
		f(\epsilon,\theta_s)&=&f_0\left(1-\beta\cos\theta_s\right)^{-
		\alpha-1}\epsilon^{-\alpha}e^{-\epsilon/E_0}\nonumber\\
		&+&f_0\left(1+\beta\cos\theta_s\right)^{-\alpha-1}
		\epsilon^{-\alpha}e^{-\epsilon/E_0},\label{eq:boost}
	\end {eqnarray}
	that is, a power law of index $\alpha$ boosted by the Lorentz factor $\Gamma=1/\sqrt{1-\beta^2}$ with an exponential cutoff energy $E_0$. The second term in
	Eq.~\ref{eq:boost} models the blazar counter jet, which we find does not significantly impact our result. To get a conservative estimate for the cascade flux,
	we employ the EBL model from \cite{Franceschini2008} which is relatively transparent for VHE gamma rays. Having no prior assumptions on the EGMF structure, we
	take $g(\theta_f)=\sin\theta_f$ to calculate the cascade flux in Eq.~\ref{eq:model}.


\section{Application of the Model to Blazar Data}\label{sec:app}
\begin {figure}[!t]
		\begin {center}
			\includegraphics[width=\columnwidth]{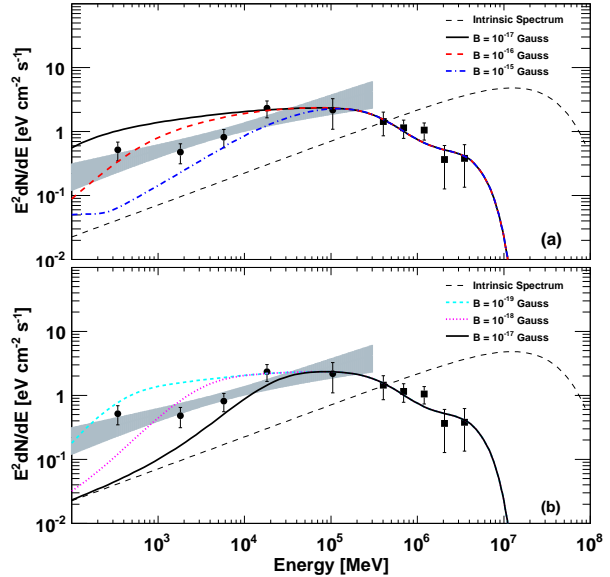}
			\caption{Measured spectrum of RGB J0710+591 (points) and predictions for different situations (curves). Squares: VERITAS data points. Shaded area: \textit{Fermi} 68\% confidence band. Circles: \textit{Fermi} data points. Panel a: Total fluxes for a source with unlimited livetime. Panel b: Total fluxes for a source with 3-year livetime.}
			\label{fig:spectrumExample}
		\end {center}
	\end {figure}

	We demonstrate the cascade model by investigating RGB J0710+591, a high-frequency-peaked BL Lacertae (HBL) object located at a redshift of
	$z=0.125$, for which simultaneous HE and VHE data are available with no variability observed~\citep{VERITAS_RGBJ0710}. In the VHE regime, we take the spectrum reported by~\cite{VERITAS_RGBJ0710}, while to get
	the HE spectrum we analyze publicly available \textit{Fermi} Large Area Telescope (LAT) data taken from a $\sim 3$-year period between 2008 August and 2011 January, using the \textit{Fermi}
	Science Tools v9r18p6 and the P6\_V3\_DIFFUSE instrument response functions (IRFs), with models gll\_iem\_v02 for the galactic diffuse and isotropic\_iem\_v02
	for the isotropic background\footnote{http://fermi.gsfc.nasa.gov/ssc/}. Accounting
	for nearby point sources, we perform an unbinned likelihood analysis to get the spectrum between 100 MeV and 300 GeV, finding a best-fit index of $1.62\pm0.11$.
	Next, we bin the data into five energy bins, demanding that each bin have a test statistic greater than 10 and a maximum relative uncertainty on the flux of $50\%$, as
	suggested by~\cite{Fermi_1FGL}. The combined spectra appear in Fig.~\ref{fig:spectrumExample}, demonstrating that the \textit{Fermi} spectral points
	are consistent with the 68\% confidence band from the unbinned likelihood analysis.

	To compare with the measured spectra, we compute the total flux as the sum of the cascade flux calculated by Eq.~\ref{eq:model} and the direct flux $f(\epsilon,0)
	\exp\left(-\tau(\epsilon)\right)$, and then normalize it to observational data. For example, the results for $\alpha=1.5$, $E_0=25$ TeV, $\Gamma=10$, and a
	range of different EGMF strengths appear in Fig.~\ref{fig:spectrumExample}(a) for a blazar with unlimited livetime, and in Fig.~\ref{fig:spectrumExample}(b)
	for a blazar active for only 3 years. In both cases we require $\theta_c$ to be smaller than the 68\% containment radius of the point spread function (PSF)~\citep{Fermi_LAT} in the \textit{Fermi} range, or
	smaller than the $\theta^2$ cut~\citep{VERITAS_RGBJ0710} in the VHE range. Setting the simple requirement that the flux be lower than the measured spectrum,
	we observe from Fig.~\ref{fig:spectrumExample} that the lower limit on the EGMF strength should be between $10^{-16}$ and $10^{-15}$ Gauss for the
	infinite-time case, and between $10^{-18}$ and $10^{-17}$ Gauss for the 3-year case, consistent with results shown by~\cite{Taylor2011} for RGB J0710+591.

	\begin {figure}[t]
		\begin {center}
			\includegraphics[width=\columnwidth]{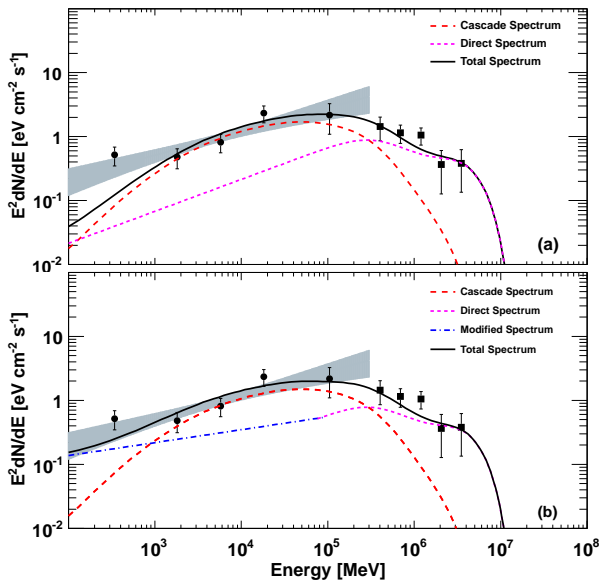}
			\caption{Fitting RGB J0710+591 spectrum with cascade model predictions assuming unlimited source livetime. Squares: VERITAS data points. Shaded area: \textit{Fermi} 68\% confidence band. Circles: \textit{Fermi} data points. Panel a: Fitting the sum of cascade flux and direct flux. Panel b: As Panel a, except the direct flux power-law index changes below 80 GeV (see text).}
			\label{fig:fitExample}
		\end {center}
	\end {figure}


	To be more rigorous, we fit the predicted flux to the data and use the resulting $\chi^2$ values to constrain the EGMF strength. Fig.~\ref{fig:fitExample}(a) shows a sample fit for $\alpha=1.5$, $E_0=25$ TeV, $\Gamma=10$, and $B=3\times 10^{-16}$ Gauss. Because we cannot exclude the existence of additional components contributing to the blazar emission~\citep[e.g.,][]{Bottcher2008} and modifying the intrinsic spectrum of Eq.~\ref{eq:boost}, we also fit a broken power law with an index $\alpha_\text{break}$ below 80 GeV, as shown for instance in Fig.~\ref{fig:fitExample}(b). We find that the additional free parameter does not greatly affect our constraints on the EGMF.

	\begin {figure}[t]
		\begin {center}
			\includegraphics[width=\columnwidth]{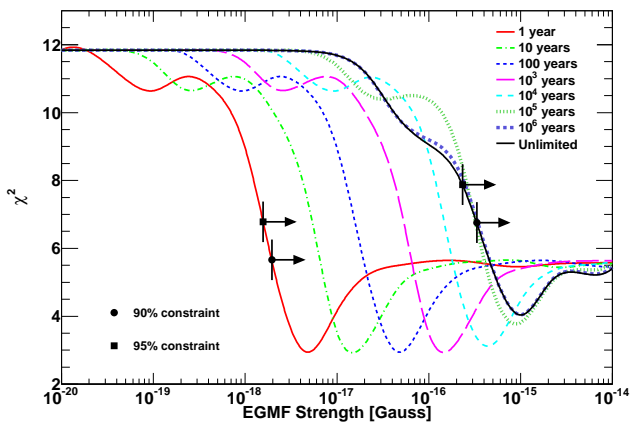}
			\caption{Best-fit $\chi^2$ versus EGMF strength $B$ for various livetimes for RGB J0710+591. Lower limits on $B$ are indicated at 90\% and 95\% confidence levels for the 1-year and unlimited cases but omitted at intermediate times for clarity.}
			\label{fig:time_cuts}
		\end {center}
	\end {figure}

	For each EGMF strength $B$, we find the best-fit model of the combined cascade and intrinsic emission for a wide range of cutoff energies
	$E_0\in(0.1$ TeV, 100 TeV), allowing $\alpha$ and $\alpha_\text{break}$ to vary between 2.5 and the physically motivated constraint of 1.5~\citep[e.g.,][]{Malkov2001,HESS_2006} and fixing $\Gamma$ to a typical value of 10. We plot the $\chi^2$ value of the best-fit model as a function of $B$ in Fig.~\ref{fig:time_cuts} for blazar livetimes from 1 year to $10^6$ years, and for the unlimited case. At low $B$, the $\chi^2$ values converge because the cascade arrives promptly and at small angles. As $B$ increases, the arrival angles and times of the cascade begin to spread out, diminishing the observed emission and providing a better fit to the observed data.

	The convergence of the curves in Fig.~\ref{fig:time_cuts} to the infinite-livetime curve is easily understood. Combining Eq.~\ref{eq:angle} with Eq.~\ref{eq:time} and assuming small angles, we can translate a cut on the instrument angle $\theta_c$ into a time cut $\Delta T_c$:
	\begin {equation}\label{eq:timecut}
		\Delta T_c\approx\frac{(L-L')L}{2cL'}{\theta_c}^2.
	\end {equation}
	For example, source photons at 1 TeV, which have a mean free path of $L'\approx400$ Mpc, will produce cascade photons of energy $E\approx0.8$ GeV, for which an angular cut of $\theta_c\approx1^\circ$ is appropriate for the \textit{Fermi} LAT. At the distance of RGB J0710+591 ($L\approx500$ Mpc), this translates into a time cut of $\Delta T_c\approx6\times10^4$ years. This is close the the livetime of $\sim10^5$ years at which the $\chi^2$ curves in Fig.~\ref{fig:time_cuts} begin to converge to the unlimited-livetime case, becoming nearly indistinguishable at $\sim10^6$ years. If the blazar livetime is smaller, the livetime cut outweighs the angle cut, and the position of the $\chi^2$ curve depends on the blazar livetime. For longer livetimes, the angle cut becomes more constraining and the curves converge to the unlimited-livetime case.

	\begin {figure}[t]
		\begin {center}
			\includegraphics[width=\columnwidth]{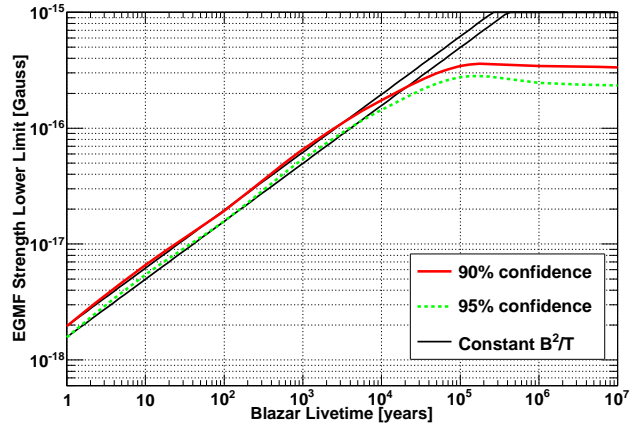}
			\caption{Excluded field strength as a function of blazar livetime for two different confidence levels for RGB J0710+591. The solid black lines of constant $B^2/T$ agree with the curves for small times but begin to diverge for livetimes between $10^4$ and $10^5$ years, as expected. At longer times, the instrument PSF limits the possible delay time of the photons, causing the constraint to level off.}
			\label{fig:confidence_limits}
		\end {center}
	\end {figure}

	We constrain the EGMF strength for a given livetime by finding the point at which $\chi^2$ exceeds its minimum value by $\Delta\chi^2$ for each curve in Fig.~\ref{fig:time_cuts}. Two sample confidence levels
	(90\% and 95\%), corresponding to $\Delta\chi^2$ values of 2.72 and 3.84~\citep[see e.g.,][]{James2006} are indicated in the figure. The limits derived from these confidence levels are shown in Fig.~\ref{fig:confidence_limits} as a
	function of the blazar livetime. For livetimes below $\sim10^4$ years, the limit on the EGMF strength scales with the blazar livetime $\Delta T$ as
	$B\sim\sqrt{\Delta T}$, as expected from combining Eq.~\ref{eq:deflection} and Eq.~\ref{eq:timecut}. This relation breaks down when the time limit enforced by
	Eq.~\ref{eq:timecut} is smaller than the blazar livetime. Our limit at 95\% confidence is $B\gtrsim2\times10^{-16}$ Gauss if the blazar's livetime is infinite
	and $B\gtrsim3\times10^{-18}$ Gauss if the livetime is 3 years, in agreement with the results of~\cite{Taylor2011}.
	In all cases, we reject the hypothesis of zero EGMF at greater than $2\sigma$ confidence.


\section{Discussion and Conclusion}\label{sec:end}
	The cascade model we present builds upon previous models to give a more complete semi-analytic treatment, accounting for the photon trajectories, background spectra, and intrinsic
	emission in sufficient detail to produce realistic cascade predictions for different blazar livetimes. For example, the spectra shown in Fig.~\ref{fig:spectrumExample}
	converge to the zero field case at high energies and become suppressed by the field approximately as $1/B$ at low energies ($\sim1$ GeV), in agreement with the spectra
	produced by a full Monte Carlo simulation as presented in \cite{Taylor2011}. Previous cascade models~\citep{Tavecchio2010,Dermer2011} overestimate the suppression
	of the cascade flux in the EGMF compared to our result.
	
	Combined with a systematic framework for interpreting the simultaneous fit of HE and VHE spectra to predictions, our model derives robust lower limits on
	the EGMF strength. The application of the model to RGB J0710+591 data yields lower limits at 95\% confidence that agree with the results of~\cite{Taylor2011}
	for the same source, further indicating that the physical picture of the cascade presented by the model is sufficiently complete to produce accurate results
	within the limits discussed in Sec.~\ref{sec:model}.
	The scaling of the lower limits with source livetime, depicted in Fig.~\ref{fig:confidence_limits}, demonstrates that our analysis framework is also physically
	meaningful.
	The limits are conservative because we choose a low EBL model, as well as for a few other reasons that we now discuss.


	We assume that the EGMF is configured as coherent domains of the same field strength and random field orientation, neglecting domain crossing by the
	$e^{\pm}$ pairs. In the energy range of interest, this restricts the domain size $\lambda_B$ to be $\lambda_B\gtrsim 1$ Mpc. If $\lambda_B\lesssim 1$ Mpc instead, domain crossing will
	result in smaller deflection angles for pairs~\citep{Ichiki2008,Neronov2009} and a higher bound on the field strength than we report.

	In testing various models of the intrinsic emission, we only consider cutoff energies below 100 TeV and spectral indices softer than 1.5. While it is
	possible to achieve intrinsic emission that is even harder by invoking some unconventional
	mechanism~\citep[e.g.,][]{Stecker2007,Bottcher2008,Aharonian2008}, the HE cascade flux could only be higher
	in that case, leading to a more stringent lower limit. The spectral
	index could also be further constrained using multi-wavelength
	information~\citep[e.g.,][]{Tavecchio2011}, which depends on detailed blazar modeling and hence
	is beyond the scope of this Letter.

	The P6\_V3\_DIFFUSE IRFs used in the cascade model integration may underestimate the PSF at large
	energies~\citep{Ando2010,Neronov2011}, but Fig.~\ref{fig:spectrumExample} shows that the most constraining part of the HE spectrum for RGB J0710+591 is
	the low-energy region where this effect is small\footnote{http://fermi.gsfc.nasa.gov/ssc/data/analysis/LAT\_caveats.html}, and, as discussed in Sec.~\ref{sec:app}, it can only affect the cases with livetimes larger than $10^4$ years. Furthermore, we expect the IRFs we use to underestimate the cascade flux,
	so our lower limit is still conservative. In the future, there will be more realistic IRFs for the \textit{Fermi} LAT, as well as more blazars with
	simultaneous HE-VHE baseline data available, and the lower limit could be further improved using the framework we present.

\acknowledgements
	The computations used in this work were performed on the Joint
	Fermilab - KICP Supercomputing Cluster, supported by grants from
	Fermilab, Kavli Institute for Cosmological Physics, and the University
	of Chicago.

	We thank V. V. Vassiliev for helpful discussions and comments in the
	preparation of this manuscript, and the anonymous referee for many
	constructive suggestions.




\bibliographystyle{apj}                       

\end{document}